\documentclass[%
 reprint,
 amsmath,amssymb,
 aps,
]{revtex4-1}

\usepackage{graphicx}% Include figure files
\usepackage{dcolumn}% Align table columns on decimal point
\usepackage{bm}% bold math
\usepackage{color}
\usepackage{xspace}

\begin{document}

\preprint{APS/123-QED}

\title{Predator-prey model for the self-organisation of stochastic oscillators in dual populations}

\author{Sara Moradi$^1$}
%\email{smoradi@ulb.ac.be}
\author{Johan Anderson$^2$}%
\author{Ozg\"{u}r. D. G\"{u}rcan$^3$}%
\affiliation{%
$^1$ Fluid and Plasma Dynamics, Universit\'{e} Libre de Bruxelles, 1050-Brussels, Belgium\\
 $^2$Department of Earth and Space Sciences, Chalmers University of Technology, SE-412 96 G\"{o}teborg, Sweden\\
  $^3$ Ecole Polytechnique, CNRS UMR7648, LPP, F-91128, Palaiseau, France}

\begin{abstract}
A predator-prey model of dual populations with stochastic oscillators is presented. A linear cross-coupling between the two populations is introduced following the coupling between the motions of a Wilberforce pendulum in two dimensions: one in the longitudinal and the other in torsional plain. Within each population a Kuramoto type competition between the phases is assumed. Thus, the synchronisation state of the whole system is controlled by these two types of competitions. The results of the numerical simulations show that by adding the linear cross-coupling interactions predator-prey oscillations between the two populations appear which results in self-regulation of the system by a transfer of synchrony between the two populations. The model represents several important features of the dynamical interplay between the drift wave and zonal flow turbulence in magnetically confined plasmas, and a novel interpretation of the coupled dynamics of drift wave-zonal flow turbulence using synchronisation of stochastic oscillator is discussed.
\end{abstract}

\pacs{Valid PACS appear here}
\maketitle

%\section{\label{sec:level1}Introduction}
\section{Introduction:}
In magnetically confined plasmas drift wave turbulence is believed to be responsible for the anomalous transport and the generation of inherent sheared zonal flows providing a self-regulating mechanism which may control the turbulence itself. The coupled system of drift wave-zonal flow (DW-ZF) is one of the key points in the evasive explanation of the low to high confinement (L-H) transition in fusion plasmas \cite{dimond2005}. The DW-ZF problem is a particular example of the more general problem of describing the nonlinear interaction and of understanding structure formation and self-organization. Such problems are ubiquitous, and notable examples are Langmuir turbulence, the dynamo problem, and the formation of ionospheric structures, just to name a few.  In such systems the self-organisation may be an example of quenched disorder being unavoidable in nature. The predator-prey paradigm \cite{lotka1920, volterra1928, goel1971, klebaner2001, mobilia2007, tauber2012} in which two subassemblies of elements (e.g. organic cells, chemical reactions, etc) influence each other through cooperation or competition has been put forward to explain this dynamical self-regulating behaviour of the dual system of DW-ZF. That is, much like a predator-prey system the damping of drift waves (i.e. prey) by zonal flows (i.e. predator) weakens the very source of zonal flow generation. There has been much progress in theoretical development of simplified models based on these ideas in the past few years \cite{dimond94,Charlton94,dimond2005,dimond2011}. Such a simplified predator-prey system constitutes of a quasi-linear Boltzmann equation for the drift wave population coupled to an equation describing the zonal flow potential growth and damping by modulational instability \cite{Rosenbluth,Hinton,miki2007}. The zonal flow potential in turn can be modified by collisional damping as well as the nonlinear damping of zonal flows such as the turbulent viscous damping. The Direct numerical simulations has played an important role in understanding the underlying mechanisms of this dual system, and have directly tested the physics of the modulational instability process \cite{tang2005}. Furthermore, several experiments have identified various elements characteristic of zonal flow phenomena, and recently limit-cycle-oscillation has been observed experimentally with the characteristics of a predator-prey system during the low-intermediate-high confinement (L-I-H) transition in various magnetically confined plasma devices \cite{conway2011, yan2014}.

In this work we developed a new family of predator-prey systems with stochastic oscillators in dual populations where the subassemblies of elements influence each other through linear and non-linear interactions. In a simplified picture we assume that the radial excursion of a drift wave eddy, and its tilting as a result of a sheared zonal flow is represented as two coupled oscillatory motions, thus in our model the two populations of the stochastic oscillators are analogous to the DW and ZF populations. The underlying reasoning is that by rewriting the function representing DW or ZF as $f_k = |f_k| exp (i \theta_k (t) + i \vec{k} \cdot \vec{r})$ and following the typical quadratically nonlinear primitive equations that arise in practice for the DW as:
\begin{eqnarray}
\frac{df_k}{dt} + i\omega_{k} f_{k} + \sum_{k=k'+k''} M_{k'k''} f_{k'} f_{k''} =0 \label{1}
\end{eqnarray}
we find a phase evolution equation of the Kuromoto form \cite{kuramoto} for each population. The Kuramoto model describing the phase dynamic of a system of stochastic limit-cycle oscillators running at arbitrary intrinsic frequencies, and coupled through the sine of their phase differences is the most successful attempt to mathematically explain the self-synchronisation phenomena \cite{kuramoto}. A system of coupled limit-cycle oscillators that can exhibit spontaneous self-organisation where the system spontaneously lock into a common frequency despite the difference in their individual natural frequencies has attracted a lot of attention for a long time. The system behaviour imitates a diversity of physical situations such as biological clocks, physiological organisms, chemical reactors \cite{winfree, kuramoto, daido1, daido2, crawford, daido3, strogatz, hong, sonnenschein, kim2003}. In these complex systems several mechanisms are at work simultaneously, and the synergy of these mechanisms results in the self-organization of the self-regulating state. Thus, the understanding of the turbulence-zonal flow problem which is one of self-organization of structures in turbulence, can help promote the understanding of self-organization processes in other systems. The idea of interpreting turbulence by stochastic oscillators, the most important reference in this context is by Kraichnan \cite{Kraichnan1961}, has offered a novel approach to capture several important features of turbulent dynamics. Similarly here, the interest in developing such a model is to understand the relation between self-organisation and self-regulation properties of turbulence and the corresponding saturation levels.

In the following we will present a reformed set of rate equations merging the Kuramoto and predator-prey paradigms and the obtained results. At the end of this paper we discuss our findings and draw conclusions.

\section{Phase coupling model}
\begin{figure}
\includegraphics[width=8cm, height=5cm]{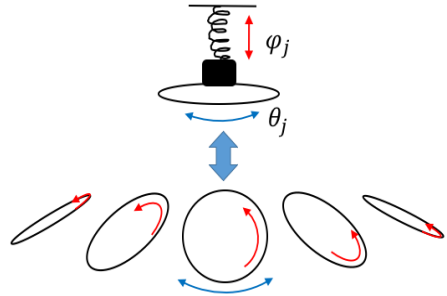}
\caption{\label{fig1} Schematics of the Wilberforce pendulum illustrate the phase variations in two dimensions where the phases $\phi$ assume to represent the fluctuating radial excursion associated to the drift wave eddies, and the phases $\theta$ assume to represent the oscillating zonal shear modulating the direction of this radial excursion via eddy tilting.}
\end{figure}

We assume an ensemble of coupled stochastic limit cycle oscillators performing a two dimensional motion similar to a Wilberforce pendulum \cite{wilberforce}. Their motion can be represented by two phases of $\theta$ in longitudinal and $\phi$ in torsional plain, respectively, see Fig. \ref{fig1} where the schematics of the Wilberforce pendulum representing the phase variations in two dimensions is shown. We further assume that the dynamics of the two phases are described by the two set of coupled first order differential equations:
\begin{eqnarray}
\dot{\theta}_{j}(t)=\omega_j+(2\pi)^{-1}\sum_{i=1}^{N}J_{ij}sin(\theta_i-\theta_j)+\frac{1}{2}\eta_L\phi_{j},\label{theta1}\\
%\nonumber\\&&+\frac{1}{2}\eta_{NL}\theta_{j}\phi_{j},\label{theta1}\\
\dot{\phi}_{j}(t)=\zeta_j+(2\pi)^{-1}\sum_{i=1}^{N}K_{ij}sin(\phi_i-\phi_j)-\frac{1}{2}\eta_L\theta_{j},\label{theta2}\\
\;\;\;(j=1,...,N).\nonumber
%\nonumber\\&&-\frac{1}{2}\eta_{NL}\phi_{j}\theta_{j}, \;\;\;(j=1,...,N).\label{theta2}
\end{eqnarray}
where we considered the $\theta_j$ and $\phi_j$ to follow a non-linear sinusoidal coupling as in the Kuramoto model \cite{kuramoto} with an additional linear cross-coupling term between the two motions. Without loss of generality hereafter we will refer to each phase equation as a population or ensemble. The eqs. (\ref{theta1}) and (\ref{theta2}) thus, represent a dual predator-prey system, where on the one hand there exists a competition between the elements within each population described by the two first terms i.e. a linear dependence given by the natural frequencies, and a non-linear dependence given by the sinusoidal function. On the other hand, there exists a competition between the elements of the two populations described by the last terms i.e. the linear cross-coupling. This cross-coupling is introduced following the coupling assumed in the Wilberforce pendulum model \cite{wilberforce} between the motions in the two dimensions which are also similar to the regular Lotka-Volterra predator-prey model \cite{lotka1920, volterra1928, goel1971}. $\dot{\theta}_{j}(t)$, and $\dot{\phi}_{j}(t)$ denote the time derivatives of the phases of $j$th oscillator in each population. The parameter $\eta_L$ measure the strength of the linear cross-coupling between the elements of two populations. $\omega_j$ and $\zeta_j$ are the natural frequencies of the oscillators in each population and for simplicity they are assumed to be both distributed according to a Gaussian distribution $f(\omega)=exp(-\omega^2/2)/\sqrt 2\pi$. $J_{ij}$ and $K_{ij}$ are the strengths of the sinusoidal interactions between oscillators $i$ and $j$ in each population, and are assumed to be random constants distributed according to an $\alpha$-stable distributions $S_{\theta,\phi}(\alpha,\beta,\sigma,\mu)$ with characteristic exponent $0<\alpha\le 2$, skewness $\beta$, scale $\sigma$ and location $\mu$ \cite{Borak,Chambers,Weron}. Here we chose $\beta=0$, $\mu=0$, and $\sigma_{\theta,\phi}=\{F,G\}/(\sqrt{2N})$ with $F$ and $G$ as control parameters as in Ref. \cite{daido2}. Moreover we assume positive and symmetric coupling, i.e. $X_{ij},>0$ and $X_{ij}=X_{ji}$ respectively. The $\alpha$-stable distributions are a general class of distributions which also include Gaussian ($\alpha=2$) and Lorentzian ($\alpha=1$) distributions. In this work, however, we focus only on the case where the coupling strengths are Gaussian distributed i.e. $\alpha=2$.

Here, we can draw an analogy to the system of DW-ZF by considering the phases $\phi$ to correspond to a fluctuating radial excursion associated to the drift wave eddies, and the phases $\theta$ to correspond to the oscillating zonal shear modulating the direction of this radial excursion via eddy tilting \cite{biglaridimondterry1990}, see the schematics shown in Fig. \ref{fig1}. To further clarify the analogy, we use the predator-prey model of the coupled system of drift wave fluctuations, $\langle N\rangle$, and zonal flow energy, $\langle U^2\rangle$, as described in Ref. \cite{dimond2005}
\begin{eqnarray}
&&(\partial_t +\gamma_{damp}) \langle U^2\rangle=  \alpha\langle U^2\rangle \langle N\rangle, \label{U2}\\
&&(\partial_t - \gamma_L + \gamma_{NL}) \langle N\rangle  =  - \alpha\langle U^2\rangle \langle N\rangle,\label{N}
\end{eqnarray}
where $\langle U^2\rangle= \sum_{qr}|U_{qr}^2|$ and $\langle N\rangle = \sum_{k}N_{k}$, with $qr$ representing the radial wavenumber of the zonal flow, and $k$ representing the mode number of the drift waves. $\gamma_{damp}$ is the nonlinear collisional damping or the saturation mechanism of zonal flow, $\gamma_L$ is the linear growth rate and $\gamma_{NL}$ is the nonlinear damping rate for the drift waves. The last terms on the rhs of eqs. (\ref{U2}) and (\ref{N}) are based on a diffusive model for the increase of the drift wave mode number $k_{\bot}$ through the random stretching of the drift waves by sheared zonal flow. Comparing the eqs. (\ref{theta1}-\ref{N}), therefore, the linear growth of drift wave instability is analogous to the natural frequencies of the oscillators, while the nonlinear damping terms are analogous to nonlinear sinusoidal coupling between the phases. The cross-coupling in eqs. (\ref{theta1}) and (\ref{theta2}) represents the cross-coupling term on the rhs of the eqs. (\ref{U2}) and (\ref{N}). The synchronisation of the phases in this analogy therefore, represents the stability of the DW-ZF system while the desynchronisation of the phases is a representation of instability of DW-ZF turbulence.

We would like to note here that we do not claim that the limit cycle oscillators is directly representative of a DW-ZF system, however, the two systems exhibit a dual predator-prey competition, and thus we are interested to examine the similarities between them.

\section{The numerical set up}
In this work, the numerical integration of eqs. (\ref{theta1}) and (\ref{theta2}) are performed using the Runge-Kutta 4th order scheme (RK4) with time stepping length $\delta t=2\pi\times dt$ where $dt$ is the optimum time interval varying for each integration while the sampling time step is $\Delta t=0.01$. The numerical integration is performed for the incoherent initial set with $\theta_j(t=0)$ and $\phi_j(t=0)$ taken to be positive gaussian distributed random values for two ensembles with $N=250$ oscillators. Here we employ an average over a number of different realisations of initial conditions and $J_{ij}, K_{ij}$ denoted by $N_{s}$, hereafter referred to as "samples". In the present study, the time span considered is of the order of $2\pi\times 5$. This time span is found to be long enough for the system to reach a steady-state and the numerical noise due to the finite size effects are absent. 

\subsection{Results of numerical simulations}
An analytic expression for the order parameter $Z(t)=\sum_{j=1}^{N} exp(i\theta_{j})/N$ was derived by Kuramoto that describes the quality of the synchronisation of the ensemble of oscillators with $0\le Z\le1$. Here, $Z=0$ corresponds to a complete a-synchronised state while $Z=1$ corresponds to a total synchronised state. In our analogy the order parameters thus, correspond to zonal flow energy and the drift wave fluctuations following $\langle U^2\rangle\rightarrow1-Z_{\theta}(t)$ and $\langle N\rangle\rightarrow1-Z_{\phi}(t)$, respectively. We have calculated the values of the order parameter separately for each populations with $\theta$ and $\phi$ phases, and averaged over $N_s$ samples denoted by $[Z_{\theta,\phi}(t)]$. In the following the results of numerical analysis for various considered cases are presented.

\begin{figure}
\includegraphics[width=8cm, height=4.6cm]{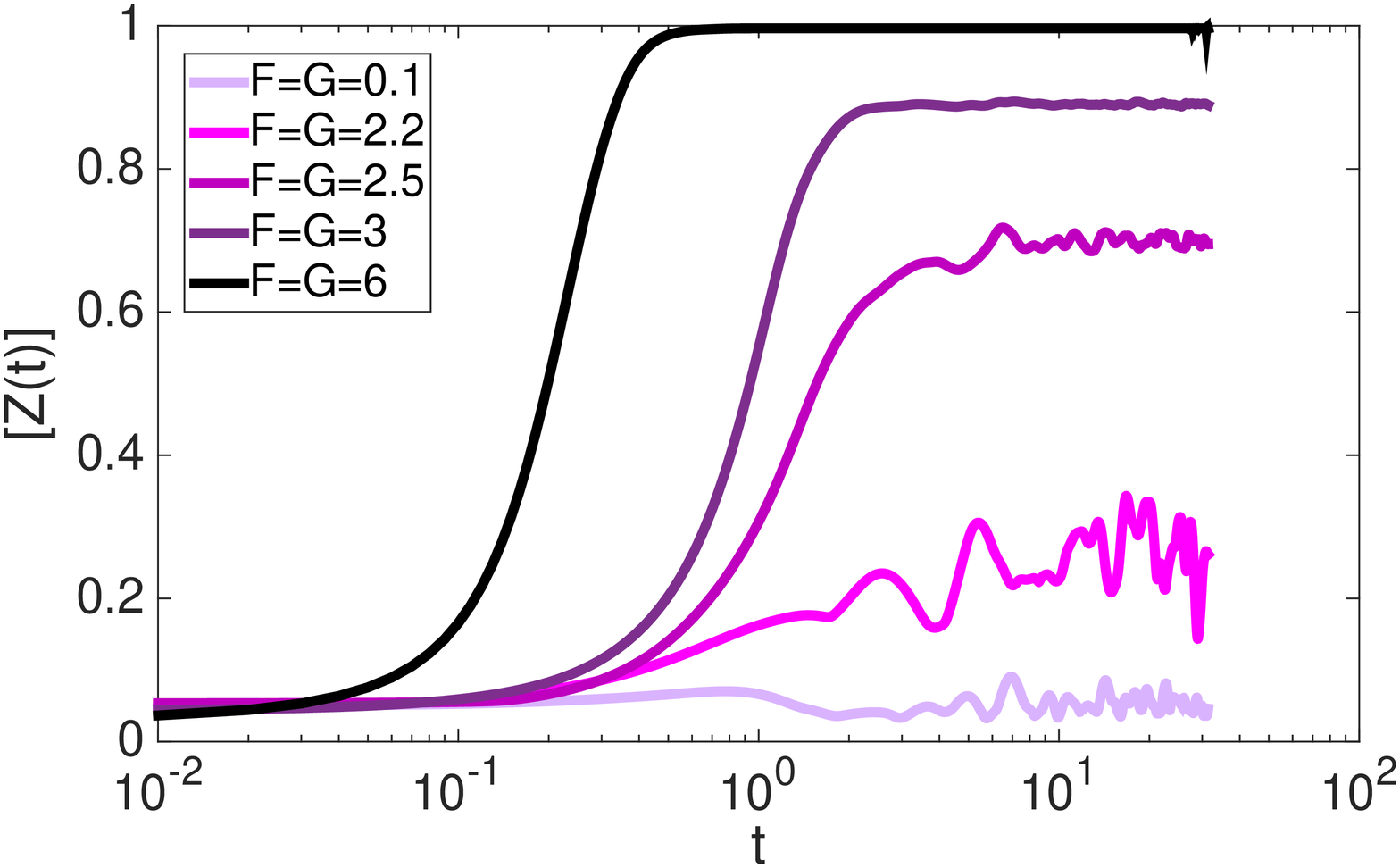}
\caption{\label{fig2a} Sample averaged $[|Z(t)|]$ as functions of time for different control parameters $F, \;G$ (note that the x-axis is logarithmic scaled). Here we have $\eta_L=0$, $F=G$, $N=250$, $N_s=5$, $\Delta t=0.01$, and the time span is $2\pi\times 5$.}
\end{figure}

\begin{figure}
\includegraphics[width=8.cm, height=4.6cm]{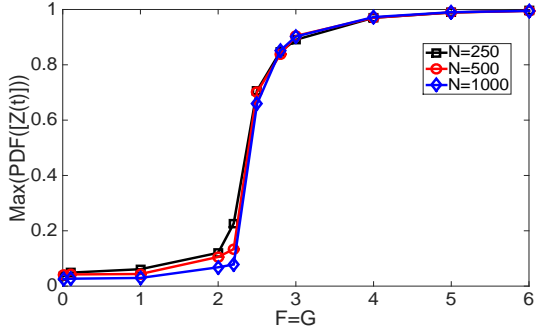}
\caption{\label{fig2b} The maximum of the $PDF([Z(t)])$ as functions of the control parameters $F, \;G$, and different number of oscillators: $N=250$ (black squares), $N=500$ (red circles) and $N=1000$ (blue diamonds). The simulation parameters are as described in the caption of Fig. \ref{fig2a}.}
\end{figure}

At first, we examine the basic model in the absence of the cross-coupling term i.e. $\eta_L=0$. Figures \ref{fig2a} and \ref{fig2b} show the evolution of the order parameter for various values of coupling strengths $F=G$ and different $N=250,\; 500$ and $1000$. In the absence of the cross-coupling term the two population follow the same behaviour. As can be seen in Fig. \ref{fig2a} for low values of control parameters i.e. $F,\;G \lesssim 2$ the phases are a-synchronised with $[Z(t)]\approx 0$. As the control parameters increase beyond this threshold the phases bifurcate from an a-synchronised to a synchronise state. The threshold where the populations change from a synchronised to a-synchronised state, in agreement with the previous reports (see Refs. \cite{daido1, daido2}), is found to be independent of the number of considered oscillators.

\begin{figure}
\includegraphics[width=8.cm, height=4.3cm]{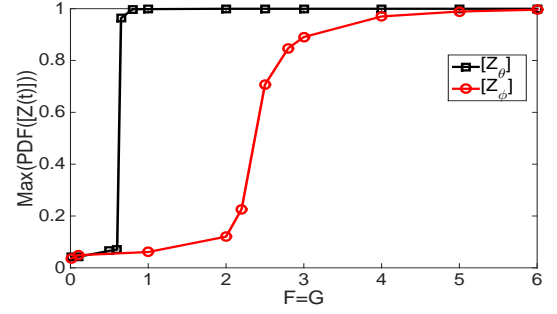}
\caption{\label{Z_vs_F_G} The maximum of the $PDF([Z(t)])$ as functions of the control parameters $F, \;G$ for $\theta$ population with $\omega=0$ (black circle symbols) and $\phi$ population with $\xi\ne 0$ (red plus symbols), for the parameters of Fig. \ref{fig2a}.}
\end{figure}

The correspondence of this trend to that of the DW-ZF is the shift from stable to unstable states as the nonlinear dampening terms are reduced, however as the two populations are not coupled to one another at this point the destabilisation of the ZF without the DW is due to the linear growth term i.e. the natural frequency $\omega$. Indeed if we neglect the natural frequency in the $\theta$ population i.e. $\omega=0$, and compare with the $\phi$ population where the natural frequencies are not neglected i.e. $\xi\ne 0$, the $\theta$ phases are found to be synchronised for a wider range of control parameter $F$, see Fig. \ref{Z_vs_F_G}. However if we decrease the sinusoidal coupling between the oscillators by decreasing $F$ the system eventually becomes desynchronised since the initial condition is an a-synchronised state and the phases are not coupled strongly enough to allow for synchronisation to take place. This transition from a synchronised to a-synchronised state occurs almost spontaneously while for the $\phi$ population further the competition between the linear term ($\xi$) and the non-linear sinusoidal term results in a more gradual transition between synchronised and a-synchronised states. The role of the natural frequencies in the phase transition of the coupled oscillators has been discussed in Refs. \cite{strogatz,Kuramoto87}. Again we would like to remind the reader that the claim here is not to make a direct comparison between the two system but to observe the similarities of the behaviours  between them. In the following computations we retain the natural frequencies of the $\theta$ population.

\begin{figure}
\includegraphics[width=4.2cm, height=5.cm]{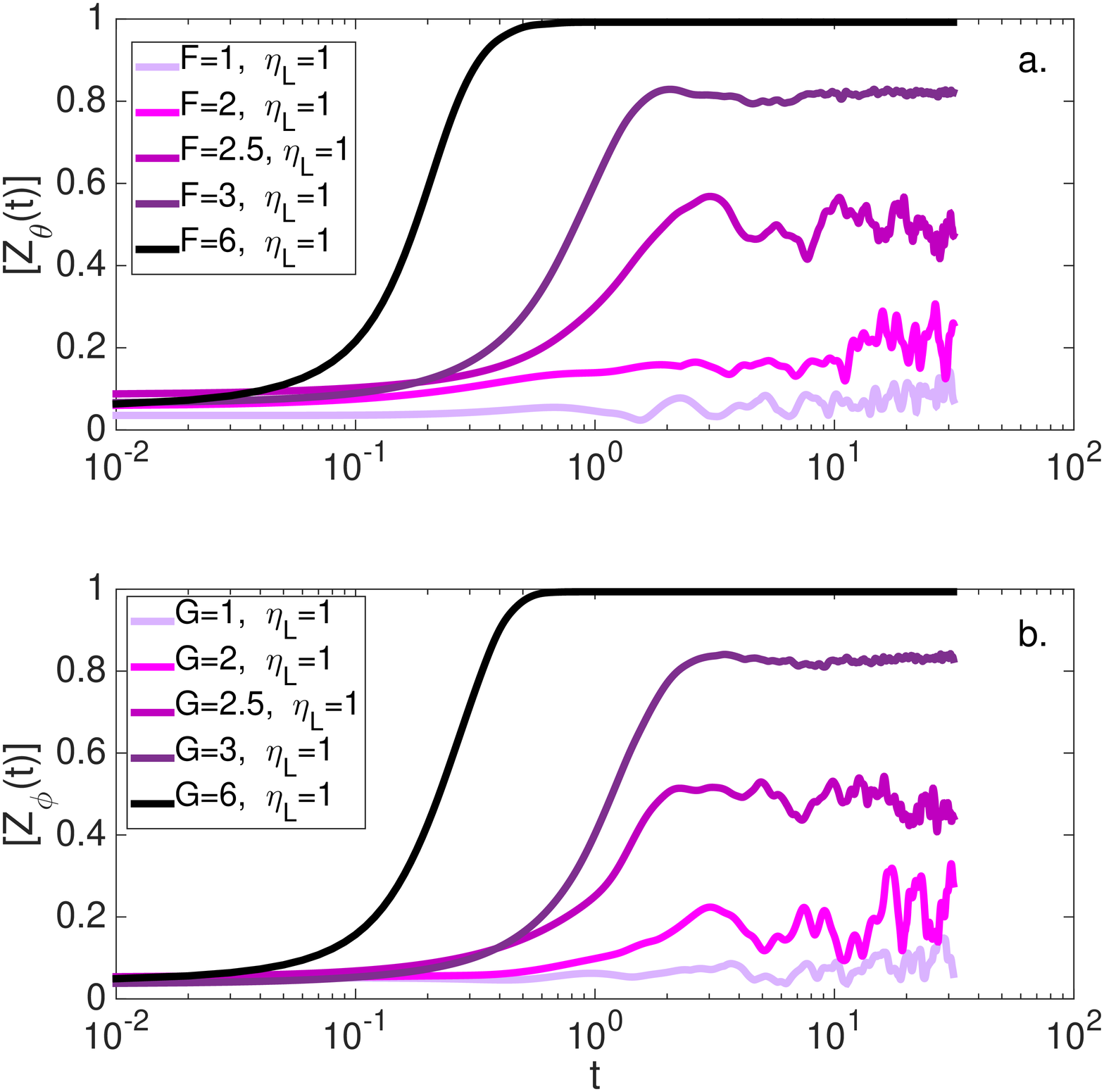}\includegraphics[width=4.2cm, height=5.cm]{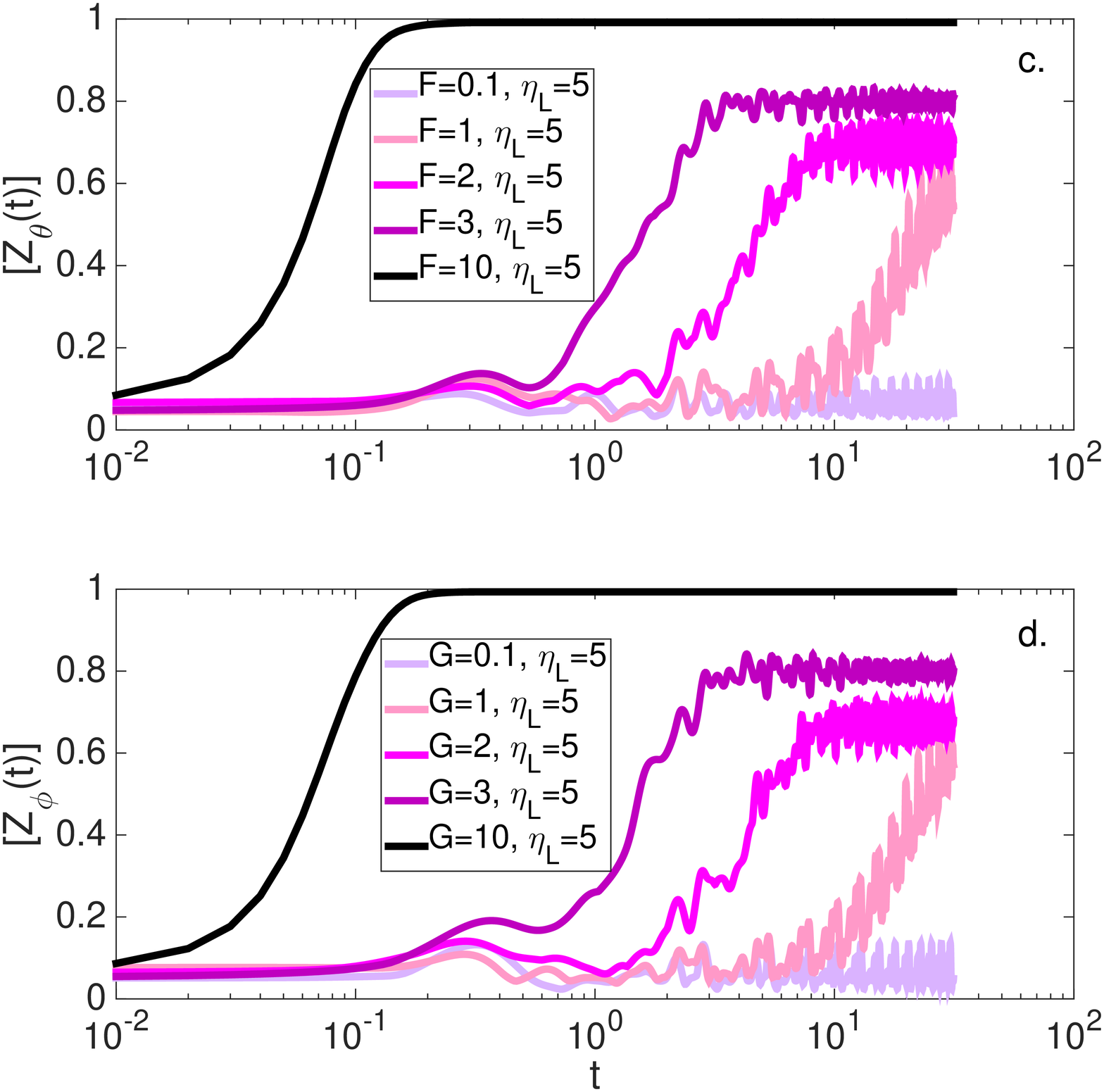}
\caption{\label{fig3ab} Sample averaged $[|Z(t)|]$ as functions of time for different control parameters $F, \;G$ for $\eta_L=1$ (a,b) and $\eta_L=5$ (c,d) (note that the x-axis is logarithmic scaled). Here we have $N=250$, $N_s=5$, $\Delta t=0.01$, and the time span is $2\pi\times 5$.}
\end{figure}

\begin{figure}
\includegraphics[width=8cm, height=4.6cm]{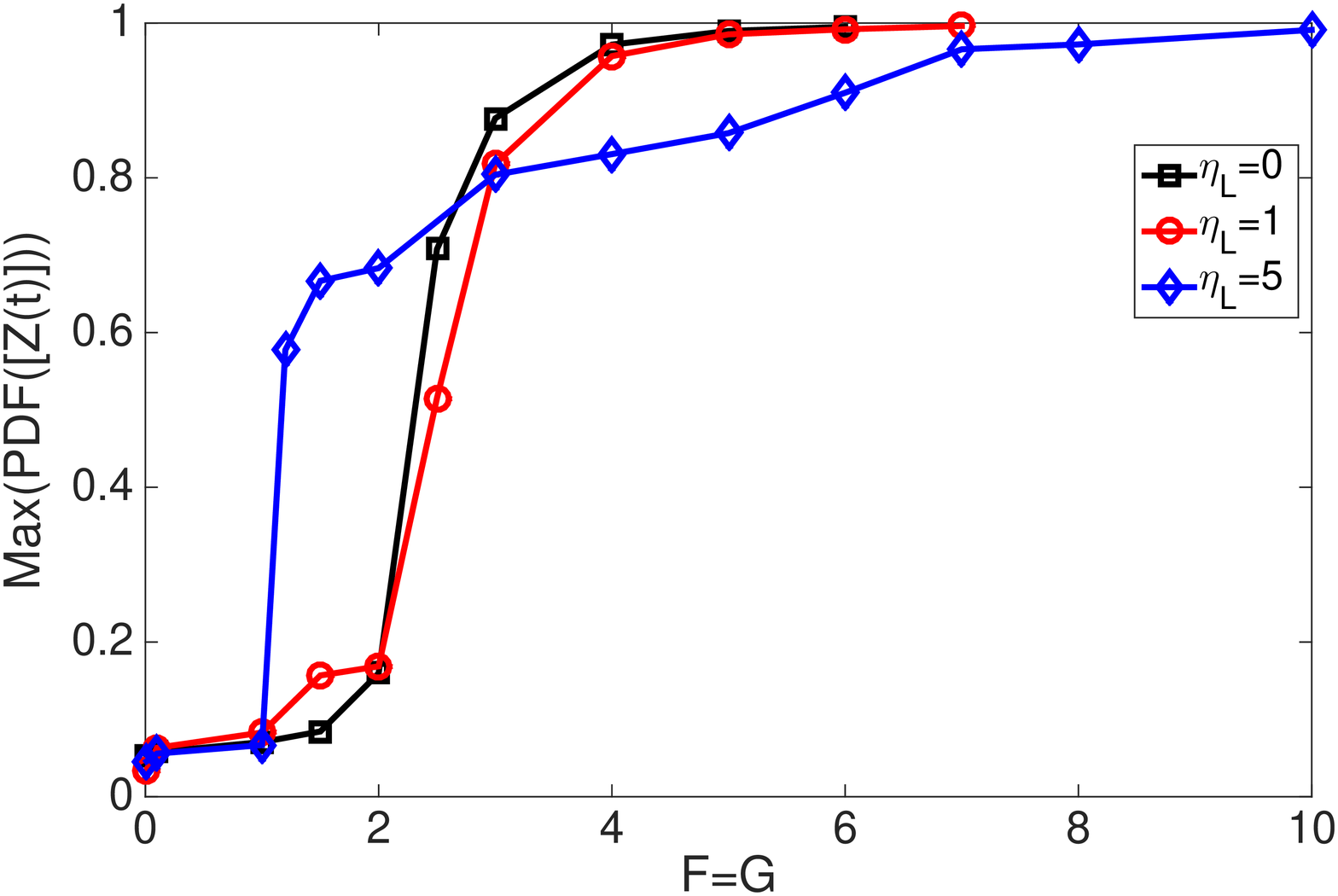}
\caption{\label{fig3e} The maximum of the $PDF([Z(t)])$ as functions of the control parameters $F, \;G$ for $\eta_L=0$ (black squares), $\eta_L=1$ (red circles), and $\eta_L=5$ (blue diamonds), for the parameters of Fig. \ref{fig3ab}.}
\end{figure}

In the next step we introduced the new feature of the model namely the linear cross-coupling. Figures \ref{fig3ab} (a-d) illustrate the time evolution of the computed order parameters for different values of $F,\;G$ with two values of linear cross-coupling strength: $\eta_L=1, 5$. Figure \ref{fig3e} shows the corresponding maximum of the $PDF([Z(t)])$ as functions of the control parameters $F,\;G$. By including a linear cross-coupling between the two populations we observe the appearance of predator-prey oscillations in the evolution of the order parameters, $[Z_{\theta,\phi}]$ with an increase in their amplitude as $\eta_L$ is increased, see Figs. \ref{fig3ab} (c and d). Another impact of the linear cross-coupling is to decrease the threshold i.e. threshold moves to lower values of $F,\; G$, where the change from an a-synchronised to a synchronised state takes place, see Fig. \ref{fig3e}. This shift of threshold is observed similarly for both populations.

Comparing to the DW-ZF system the addition of cross-coupling in the above system is similar to adding the nonlinear term on the rhs of eqs. (\ref{U2}) and (\ref{N}) which results in stabilising the DW through sheared ZF. The oscillations observed here also show similarities to that of the modulational instability observed in the DW-ZF system \cite{dimond2005,miki2007,Rosenbluth,Hinton}.

\begin{figure}
\includegraphics[width=8cm, height=7.5cm]{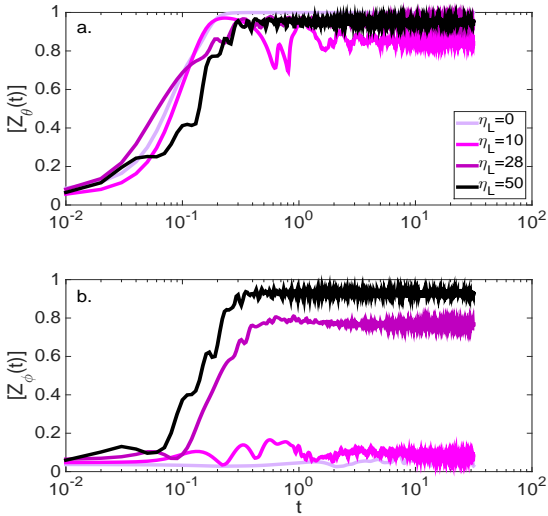}
\caption{\label{fig4ab} (a and b) Sample averaged $[|Z(t)|]$ as functions of time for different linear cross-coupling parameters $\eta_L$ with $F=10$, and $G=0.1$ (note that the x-axis is logarithmic scaled). Here we have $N=250$, $N_s=5$, $\Delta t=0.01$, and the time span is $2\pi\times 5$.}
\end{figure}

\begin{figure}
\includegraphics[width=8cm, height=4.6cm]{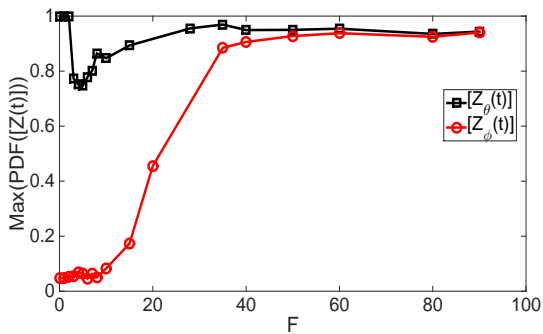}
\caption{\label{fig4c}  The maximum of the $PDF([Z(t)])$ as functions of the linear cross-coupling parameters $\eta_L$ for $\theta$ (black squares) and $\phi$ (red circles) populations, for the parameters of Fig. \ref{fig4ab}.}
\end{figure}

To investigate the importance of the cross-coupling between the two populations, starting from zero we slowly increase $\eta_L$ for a case in which $\theta$ population is synchronised with $[Z_{\theta}]=1$, by setting $F=10$, while the $\phi$ population is a-synchronised ($[Z_{\phi}]\sim 0$) by setting $G=0.1$. The reason for this choice of parameters is to mimic the situation where the nonlinear dampening of the ZF is high and therefore they are stable, while the DW are strongly unstable. Figures \ref{fig4ab} (a and b) and \ref{fig4c} show the results of this investigation. We find that initially the increase in $\eta_L$ allows for the a-synchronisation of the $\phi$ population to affect the synchronisation state of the $\theta$ population and thereby reducing $[Z_{\theta}]$. However, beyond a critical $\eta_L\sim15$ the synchrony of the $\theta$ overcomes the desynchronising effects of the $\phi$ population and even pushes them towards syntonisation. Thus, the whole system self-regulates itself where the synchronisation of one population is transferred to the other, and both systems converge to a new level of partially synchronised state with $[Z(t)]\sim 0.95$. This behaviour shows similarities to the predator-prey model of ZF generation by DW turbulence with the back reaction of the ZF on the DW and thereby stabilisation of DW. 

\begin{figure}
\includegraphics[width=8cm, height=7.5cm]{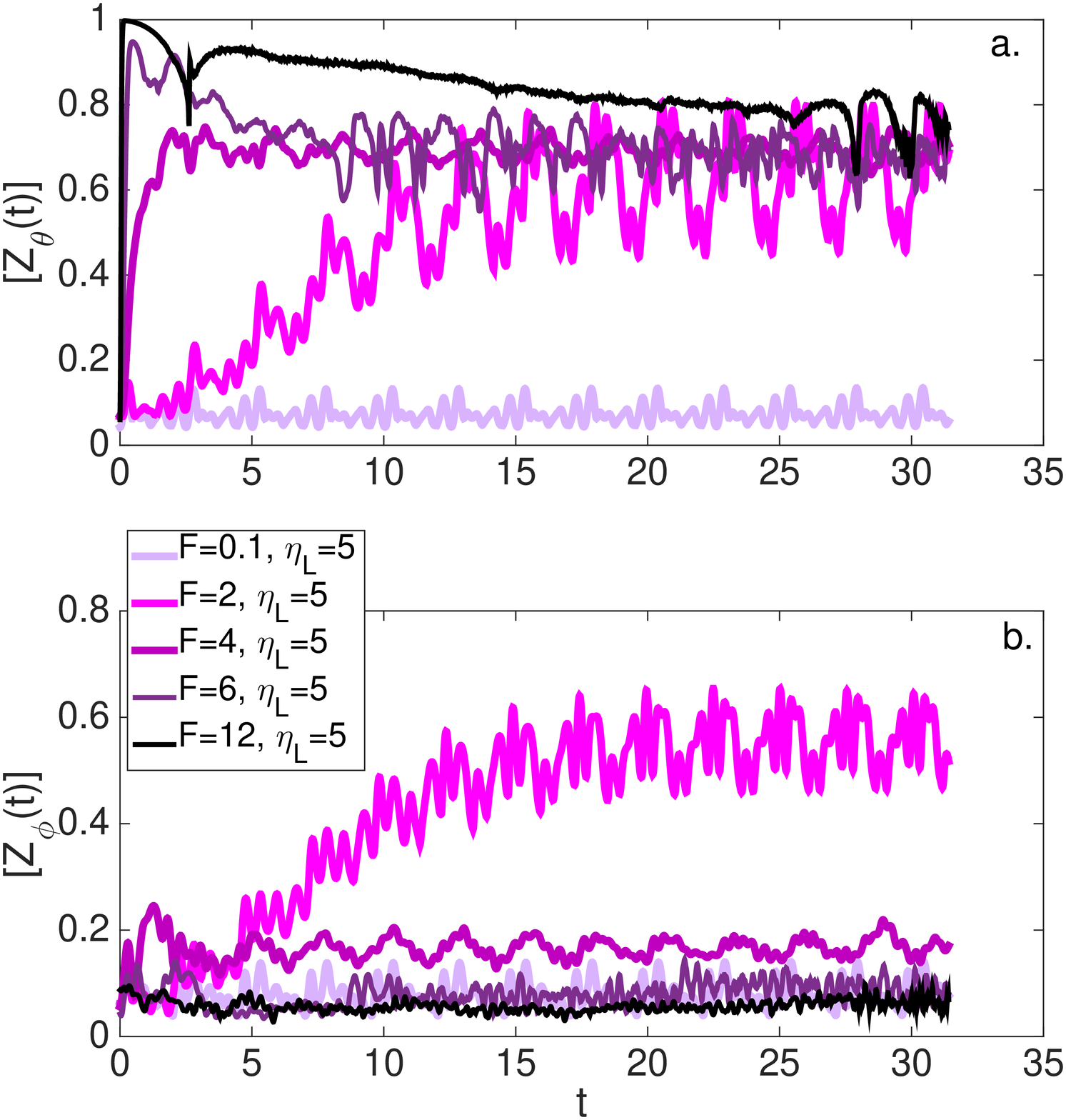}
\caption{\label{fig6ab} Sample averaged $[|Z(t)|]$ as functions of time for different control parameter $F$ for $\eta_{L}=5$ (a,b) (note that the x-axis is logarithmic scaled). Here we have $N=250$, $N_s=5$, $\Delta t=0.01$, and the time span is $2\pi\times 5$.}
\end{figure}

\begin{figure}
\includegraphics[width=8cm, height=4.3cm]{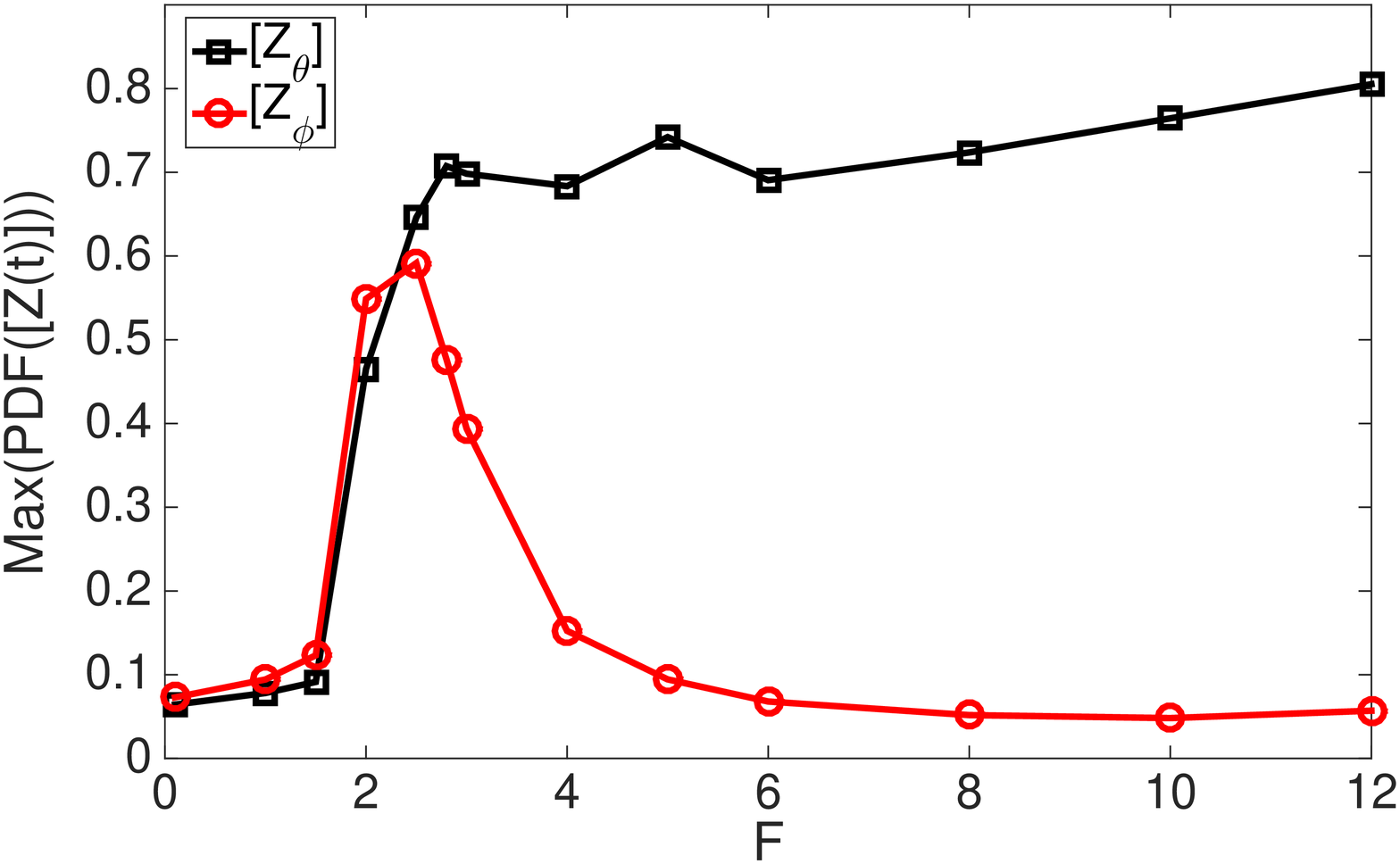}
\caption{\label{fig6c} The maximum of the $PDF([Z(t)])$ as functions of the control parameter $F$ for $\eta_{L}=5$ for $\theta$ (black squares) and $\phi$  populations (red circles), computed with the parameters of Fig. \ref{fig6ab}.}
\end{figure}

\begin{figure}
\includegraphics[width=8cm, height=7.5cm]{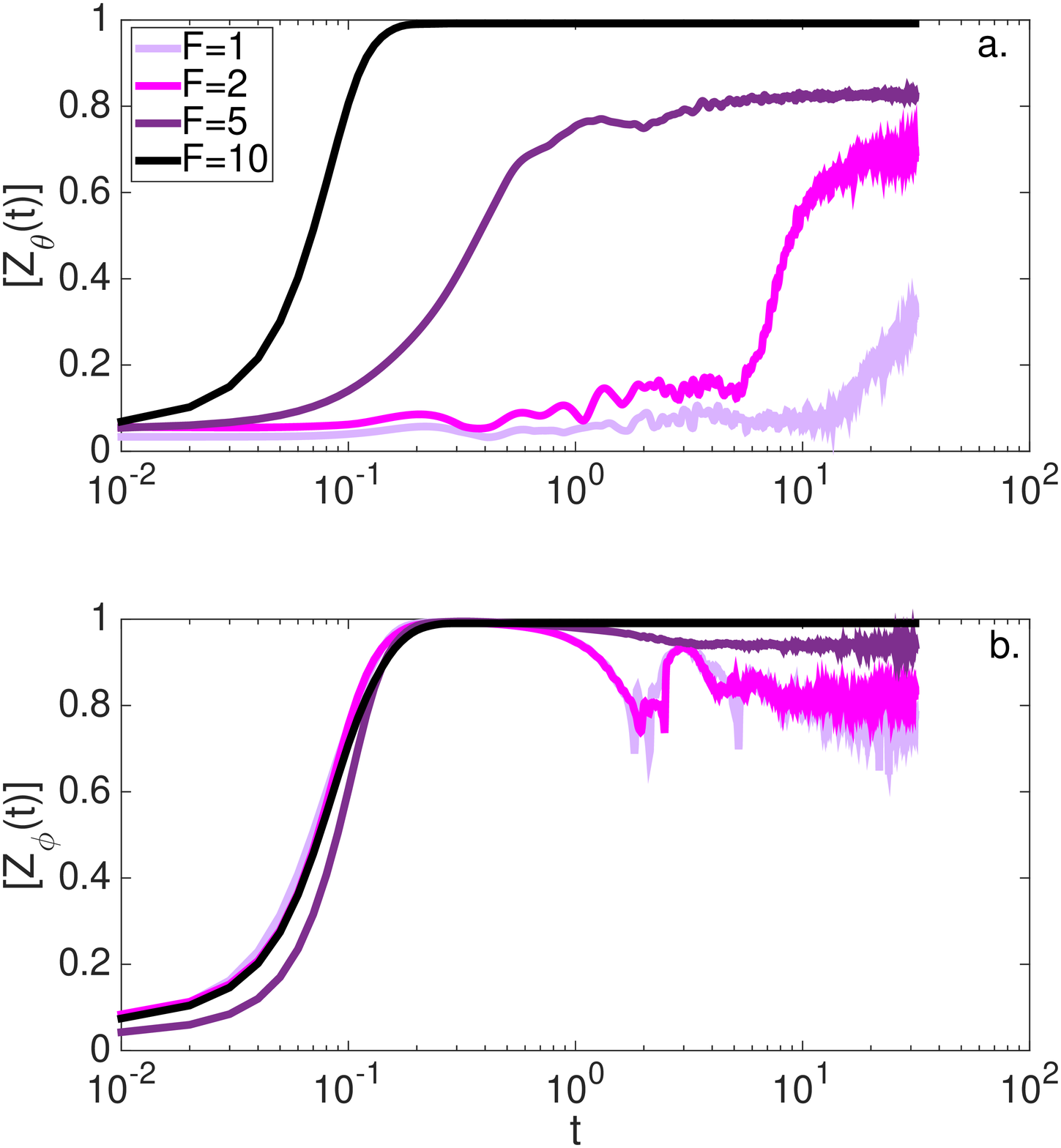}
\caption{\label{fig7ab} Sample averaged $[|Z(t)|]$ as functions of time for different control parameter $F$ for $\eta_{L}=5$ (a,b) (note that the x-axis is logarithmic scaled). Here we have $N=250$, $N_s=5$, $\Delta t=0.01$, and the time span is $2\pi\times 5$.}
\end{figure}

\begin{figure}
\includegraphics[width=8cm, height=4.3cm]{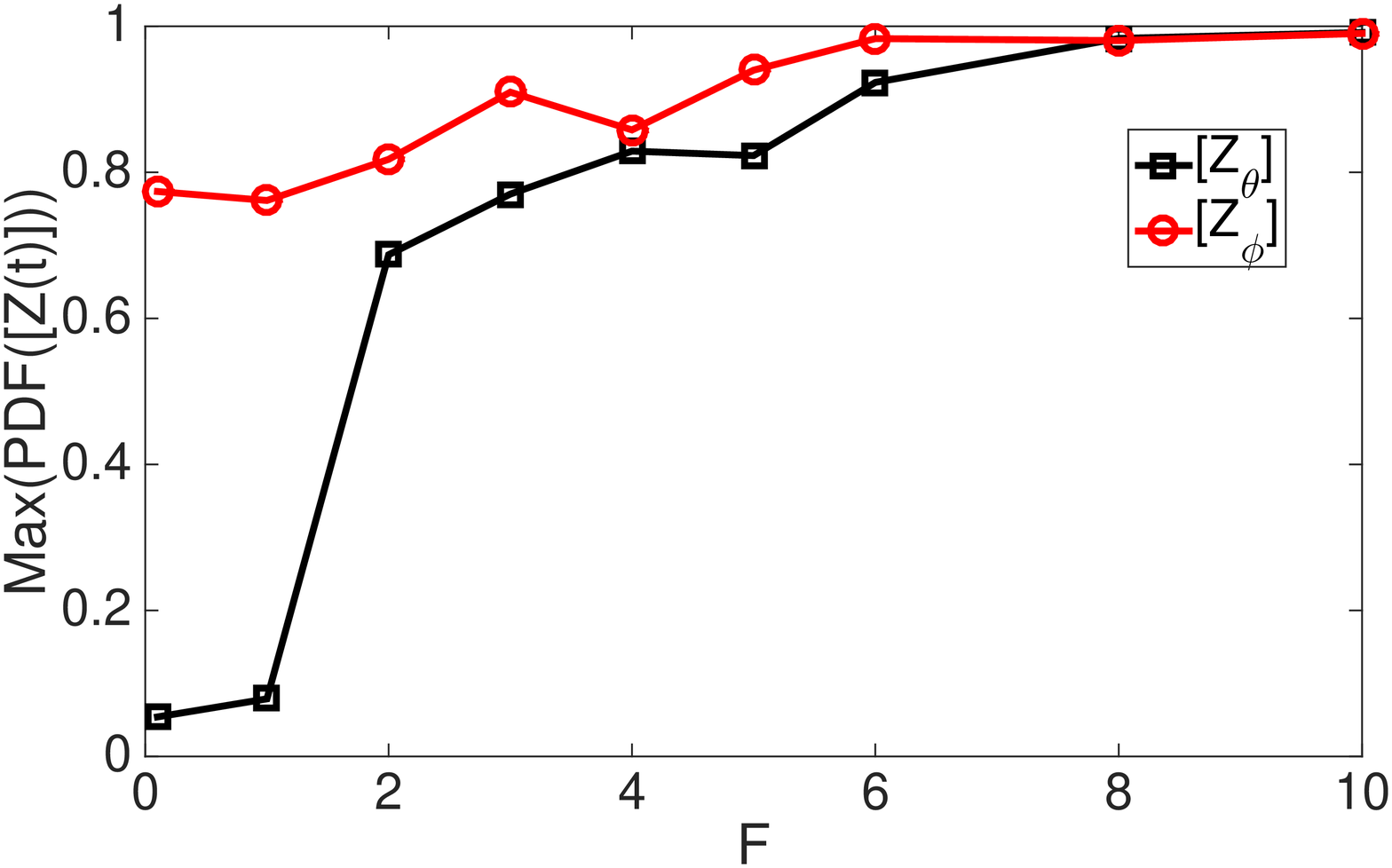}
\caption{\label{fig7c} The maximum of the $PDF([Z(t)])$ as functions of the control parameter $F$ for $\eta_{L}=5$ for $\theta$ (black squares) and $\phi$  populations (red circles), computed with the parameters of Fig. \ref{fig6ab}.}
\end{figure}

Further, we examine the relative importance of the sinusoidal versus linear cross-coupling terms, by varying the control parameter $F$, while setting $G=0.1, 10$, and $\eta_L=5$. This choice of parameters is motivated to mimic the interplay between the DW-ZF system in two situations where (i) the DW are strongly unstable, (ii) the DW are weakly unstable and the ZF dampening, $\gamma_{damp}$, varies from high (high $F$) to low (low $F$) values. The results are shown in Figs. \ref{fig6ab}-\ref{fig7c} where the evolution of the sample averaged $[|Z_{\theta,\phi}(t)|]$ and their corresponding maximum of the $PDF([Z(t)])$ for different values of control parameter $F$ are shown. The decrease in the sinusoidal coupling between the $\theta$ phases results in their desynchronisation, however, there exist a significant difference in the desynchronisation rate as a function of $F$ between the two considered cases (i) and (ii). As can be seen in Figs. \ref{fig6ab} and \ref{fig6c}, for the case (i) where the $\phi$ population is a-synchronised ($G=0.1$), the increase in $F$ results in synchronisation of $\theta$, however the system will only reach a partial synchronised state with $[Z_{\theta}(t)]\approx 0.8$ as $F$ is further increased. At the same time with increasing $F$ the $\phi$ population moves from a-synchronised state towards a partial synchronisation with $[Z_{\theta}(t)]\approx 0.6$ at $F\sim3$. By further increasing $F$ beyond this value, the sinusoidal coupling between the phases in $\theta$ population is too strong as compared to the linear cross-coupling between the two populations, and thus the coupling between the two system is diminished and the $\phi$ population returns to its a-synchronised state. The amplitude of the oscillations seen on the evolution of the order parameters increase when the coupling between the two system is at its maximum i.e. $F\sim 2-3$. This behaviour can be compared to the destabilisation of strongly damped ZF by strong DW turbulence and the stabilisation of DW by the ZF. However if the ZF dampening rate, $\gamma_{damp}$, is very low and DW are strongly unstable they can coexist due to tertiary instability i.e. the modulational instability \cite{dimond2005,miki2007,Rosenbluth,Hinton}.
The situation is different when the $\phi$ population is synchronised, $G=10$, and $[Z_{\phi}]=1$. As seen in Figs. \ref{fig7ab} and \ref{fig7c} in the presence of a cross-coupling between the two populations, the transition of the $\theta$ system from an a-synchronised to a synchronised state occurs more gradually as compared to the case without the cross-coupling, see Fig. \ref{fig2b}. At low values of $F$ where the $\theta$ population is a-synchronised or partially synchronised, a slight reduction of the order parameter for $\phi$ population i.e. a slight desynchronisation is observed which converges to $[Z_{\phi}]\sim 0.8$. These results indicated that in the case of a strong sinusoidal coupling between $\phi$ population, in order for the $\theta$ population to impose its desynchronised state on the $\phi$ population, the cross-coupling strength has to be stronger. To compare with the ZF-DW; in the absence of strong DW instability, the ZF generation by DW is limited and in a sense the coupling between them will be weak. In such a situation the reduction of the damping terms in the ZF equation will allow for their coexistence with near stability DW.
\section{Summary} 
In this work we have developed a novel model of stochastic oscillators obeying predator-prey rate equations to mimic the coupled dynamics of DW-ZF turbulence by which the system regulates and organises itself. The model assumes that the dual populations represent the oscillations analogous to the two dimensional motion of a Wilberforce pendulum: one in the longitudinal and the other in torsional plain. Within each population a Kuramoto type competition between the phases is assumed with an additional linear cross-coupling between the dual populations. Thus, the synchronisation state of the whole system is controlled by two types of competition. The results of the numerical simulations show that if the two populations are linearly cross coupled the system undergoes a modulational synchrony transfer between the two populations similar to predator-prey oscillations in DW-ZF system. Several important features of the DW-ZF system were tested against the presented model which show strong similarities. Note that the nonlinear terms in the presented model is representative of a quadratic nonlinearity. Furthermore, the dynamics in the DW-ZF is determined by non-linear terms of second order with scale dependent coefficients. This complication is at present neglected however the qualitative behaviour of the DW-ZF predator prey is captured, hence, we believe that this work offers a valid interpretation of the dynamical system with a natural mechanism for saturation through the synchronisation of stochastic oscillators.

\section{Acknowledgments}
Sara Moradi has benefited from a mobility grant funded by the Belgian Federal Science Policy Office and the MSCA of the European Commission (FP7-PEOPLE-COFUND-2008 n$¼$ 246540).


\begin{thebibliography}{100}
\bibitem{dimond2005} P. H. Diamond, S-I Itoh, K. Itoh and T. S. Hahm {\em Plasma Phys. Control. Fusion} {\bf 47} R35ÐR161 (2005).
\bibitem{lotka1920} A. J. Lotka, {\em Proc. Natl. Acad. Sci U.S.A.} {\bf 6} 410 (1920); {\em J. Amer. Chem. Soc.} {\bf 42} (1920). 
\bibitem{volterra1928} V. Volterra, {\em Mem. Acad. Lincei Roma} {\bf 2} 31 (1926).
\bibitem{goel1971} N. S. Goel, S. C. Maitra and E. W. Montroll, {\em Rev. Mod. Phys.} {\bf 43} 231 (1971).
\bibitem{klebaner2001} F. C. Klebaner and R. Lipster, {\em Ann. Appl. Probability} {\bf 11} 1263-1291 (2001).
\bibitem{mobilia2007} M. Mobilia, I. T. Georgiev and U. C. T\"{a}uber, {\em J. Stat. Phys.} {\bf 128} 447 (2007).
\bibitem{tauber2012} U. C. T\"{a}uber, {\em J. Phys. A Math. and Theor.} {\bf 45} 405002 (2012).
\bibitem{dimond94} P. H. Diamond, Y-M Liang, B. A. Carreras and P.W. Terry, {\em Phys. Rev. Lett.} {\bf 72} 2565 (1994)
\bibitem{Charlton94} L. A. Charlton, B. A. Carreras, V. E. Lynch, K. L. Sidikman and P. H. Diamond, {\em Phys. Plasmas} {\bf 1} 2700 (1994)
\bibitem{dimond2011} P. H. Diamond, A. Hasegawa and K. Mima, {\em Plasma Phys. Control. Fusion} {\bf 53} 124001 (2011).
\bibitem{Rosenbluth} M. N. Rosenbluth and F. L. Hinton, {\em Phys. Rev. Lett.} {\bf80} 724 (1998).
\bibitem{Hinton} F. L. Hinton and M. N. Rosenbluth, {\em  Plasmas Phys. Control. Fusion} {\bf 41} A653 (1999).
\bibitem{miki2007} K. Miki, Y. Kishimoto, N. Miyato and J. Q. Li, {\em Phys. Rev. Lett.} {\bf 99} 145003 (2007).
\bibitem{tang2005} W. M. Tang and V. S. Chan, {\em Plasma Phys. Control. Fusion} {\bf 47} R1 (2005).
\bibitem{conway2011} G. D. Conway, C. Angioni, F. Ryter, P. Sauter, and J. Vicente, {\em Phys. Rev. Lett.} {\bf 106} 065001 (2011).
\bibitem{yan2014} Z. Yan, G. R. McKee, R. Fonck, P. Gohil, R. J. Groebner, and T. H. Osborne, {\em Phys. Rev. Lett.} {\bf 112} 125002 (2014).
\bibitem{kuramoto} Y. Kuramoto, Chemical Oscillations, Waves and Turbulence, {\em Springer, Berlin} (1984).
\bibitem{winfree} A. T. Winfree, The Geometry of Biological Time, {\em Springer, New York} (1980).
\bibitem{daido1} H. Daido, {\em Prog. Theor. Phys.} {\bf 77} 622 (1987).
\bibitem{daido2} H. Daido, {\em Phys. Rev. Lett.} {\bf 68} 1073 (1992).
\bibitem{crawford} J. D. Crawford, {\em J. Stat. Phys.} {\bf 74} 1047 (1994).
\bibitem{daido3} H. Daido, {\em Phys. Rev. E} {\bf 61} 2145 (2000).
\bibitem{strogatz} S. H. Strogatz, {\em Physica D} {\bf 143} pp 1-20 (2000).
\bibitem{hong} H. Hong and S. H. Strogatz, {\em Phys. Rev. Lett.} {\bf 106} 054102 (2011).
\bibitem{sonnenschein} B. Sonnenschein and L. Schimansky-Geier, {\em Phys. Rev. E} {\bf 88} 052111 (2013).
\bibitem{kim2003} E. Kim and P. Diamond, {\em Phys. Plasmas} {\bf 10} 1698 (2003).
\bibitem{Kraichnan1961} R. H. Kraichnan, {\em J. Math. Physics} {\bf 2} 124 (1961).
\bibitem{wilberforce} R. E. Berg and T. S. Marshall, {\em Am. J. Phys.} {\bf 59} 32 (1991). 
\bibitem{Borak} Sz. Borak, W. H\"{a}rdle and R. Weron, {\em Statistical Tools for Finance and Insurance, eds.} Springer-Verlag, Berlin, 21-44 (2005), http://ideas.repec.org/p/hum/wpaper/sfb649dp2005-008.html.
\bibitem{Chambers} J. M. Chambers, C. L. Mallows and B. W. Stuck, {\em A Method for Simulating Stable Random Variables}, JASA {\bf 71}, 340-344 (1976).
\bibitem{Weron} R. Weron, J. E. Gentle, W. H\"{a}rdle and Y. Mori, {\em Computationally intensive Value at Risk calculations in Handbook of Computational Statistics: Concepts and Methods, eds.} Springer, Berlin, 911-950 (2004).
\bibitem{biglaridimondterry1990} H. Biglari, P. H. Diamond and P. W. Terry, {\em Phys. Fluids} B {\bf 2} 1 (1990).
\bibitem{Kuramoto87} Y. Kuramoto, I. Nishikawa, {\em Journal of Statistical Physics}, {\bf 49} 569-605 (1987).
\end{thebibliography}
\end{document}